\documentclass[twocolumn,aps,prb,notitlepage,10pt]{revtex4-1}
\usepackage{graphicx,amsmath,bm,amssymb,color}
\usepackage[colorlinks=true,linkcolor=blue,citecolor=blue]{hyperref}

\begin{document}
\title{Computing energy barriers for rare events from hybrid quantum/classical \\
  simulations through the virtual work principle}
\author{Thomas D Swinburne}
\altaffiliation{Current Address: Theoretical Division T-1, Los Alamos National Laboratory, Los Alamos, New Mexico, 87545, USA}
\affiliation{CCFE, Culham Science Centre, Abingdon, Oxon, OX14 3DB, UK}
\author{James R Kermode}
\affiliation{Warwick Centre for Predictive Modelling, School of Engineering, University of
Warwick, Coventry CV4 7AL, UK}
\date{\today}
\begin{abstract}
    Hybrid quantum/classical techniques can flexibly couple {\it ab initio} simulations to an empirical or elastic medium to model materials systems that cannot be contained in small periodic supercells. However, due to electronic non-locality a total energy cannot be defined, meaning energy barriers cannot be calculated. We provide a general solution using the principle of virtual work in a modified nudged elastic band algorithm. Our method enables the first {\it ab initio} calculations of the kink formation energy for $\langle100\rangle$ edge dislocations in molybdenum and lattice trapping barriers to brittle fracture in silicon.
\end{abstract}
\maketitle

\section{Introduction}

The two-way chemomechanical coupling of chemical and elastic fields
creates inextricably multiscale problems with a simultaneous requirement for
chemical accuracy and large system sizes. Density functional theory (DFT) has
been shown to have excellent predictive power\cite{Burke2012}, but its typically high $O(N^3)$
computational cost limits its application to problems with fewer than around 1000 atoms\cite{Martin}. This problem is particularly acute for crystal defects such as dislocation lines\cite{woodward2002}, grain boundaries\cite{wright1994} and cracks\cite{kermode2008}, which all posses a long range elastic field that can rarely be contained in small periodic supercells without unrealistically strong image interactions or strain gradients. Whilst linear elastic corrections have successfully removed finite size effects for small point defect clusters\cite{clouet2013} and screw dislocation dipoles in bcc metals\cite{Dezerald2014}, in the majority of cases crystal defects require very large supercells which even $O(N)$ first principles approaches\cite{Hine2009} cannot readily accommodate, especially in metallic systems. Furthermore, complex processes such as dislocation emission or thermally activated crack growth occur on timescales that are far too slow for direct dynamical simulations at the {\it ab initio} level. As a result it is necessary to determine rare event rates using transition state theory\cite{kramers}, for which the ability to calculate energy barriers is essential, using, e.g. the nudged elastic band (NEB) method\cite{neb}.

Large systems can be accurately modelled by combining a local QM
description with classical models using hybrid multiscale approaches~\cite{Bernstein2009} such as
the quantum mechanics/molecular mechanics (QM/MM)~\cite{Lu2006} and
`Learn on the Fly' (LOTF) schemes~\cite{Csanyi2004}. The LOTF approach has been used extensively to perform {\it ab initio} molecular dynamics\cite{kermode2008} but is limited to dynamical simulations and cannot compute energy barriers. Energy-based QM/MM schemes for metals developed by Gang Lu and coworkers~\cite{Lu2006} have been applied to energy pathways~\cite{Zhang2010,Zhang2010b}, but do not provide seamless coupling for materials systems\cite{Bernstein2009} and  require the definition of system-specific interaction potentials between the QM and MM regions, restricting the generality of the approach.

Flexible boundary DFT calculations couple a fully quantum mechanical simulation to an infinite continuum through a lattice Green's function (LGF)~\cite{woodward2002,Woodward2005,Woodward2008,Trinkle2008,Tan2016}. These methods are ideally suited to crystal defect calculations, as the heavily deformed defect core is treated quantum mechanically whilst the weakly deformed elastic field is captured in the bulk region for comparitively negligible computational cost~\cite{rodney2017}. However, while elastic embedding methods allow complex local chemical effects to be modelled they cannot include thermal or entropic effects and rely on the existence of analytical elastic solutions not readily available for complex three dimensional problems. Recent work to numerically compute the lattice Green's function of large scale defects extends the applicability of the approach~\cite{Tan2016}, but it remains restricted to structural optimisation and does not yet allow energy barriers or temperature effects to be modelled. Similarly, the QM-CADD approach~\cite{Nair2010} couples a DFT region directly with a finite element model but also cannot be used to compute energy barriers.

While hybrid and flexible boundary DFT calculations have been successfully applied to treat a wide range of crystal defects, they suffer from a well known limitation - due to the non-locality of the electronic energy an `energy-per-ion' cannot be defined in the quantum mechanical (QM) region, meaning that the total system energy, which in principle should be a sum of classical and quantum contributions, cannot be defined\cite{rodney2017}. (We note that energy differences could in principle be calculated by relying on cancellation of errors near boundaries, however, this uncontrolled assumption would have to be tested on a case-by-case basis.) As a result, important and highly desirable quantities such as migration barriers and segregation energies have long been considered inaccessible.

In this paper we detail a general solution to the problem of extracting energy barriers from hybrid simulation schemes without a total energy function. We exploit the fact that ionic forces in both the classical and quantum region are well defined and localised, allowing us to apply the principle of virtual work to construct energy barriers for a given configurational pathway. Combining this principle with the nudged elastic band routine for finding minimum energy pathways allows the calculation of energy barriers in systems much larger than can be treated in periodic DFT supercells. This is related to using thermodynamic integration to reconstruct free energy profiles in biochemical QM/MM methods~\cite{Hu2008,Varnai2013}, with the key difference that here we target zero temperature potential energies since entropic effects are comparitively small in hard condensed matter systems. We demonstrate our method on two problems typically considered inaccessible to {\it ab initio} methods, kink formation on $\langle100\rangle$ edge dislocations in Mo and lattice trapping barriers to brittle fracture in Si.

\section{Hybrid Simulation Scheme}

\begin{figure}
    \includegraphics[width=\columnwidth]{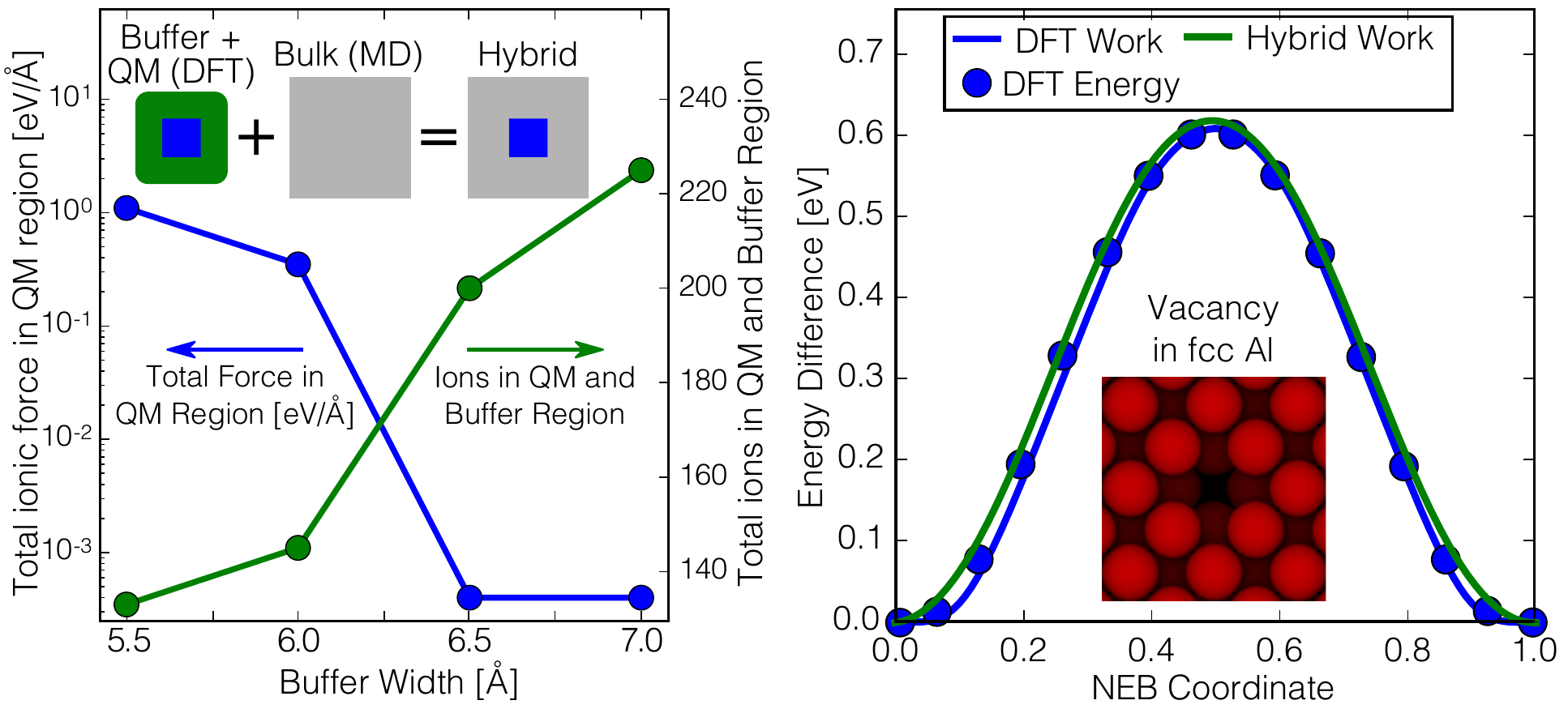}
    \caption{Left: Convergence of DFT forces in a QM region of 12 perfect aluminium lattice ions with increasing buffer width. Inset: Cartoon of the hybrid system. Right: Comparison of NEB methods to calculate the migration barrier of a vacancy in fcc aluminium.}
    \label{vacancy}
\end{figure}

A prototypical hybrid simulation scheme is shown in Fig.~\ref{vacancy}. To provide correct forces on atoms in the QM region, at each force call a DFT calculation is performed which contains the QM region, a surrounding `buffer' region and a vacuum layer to remove periodic image effects. The presence of free surfaces in the DFT supercell induces electronic (though not elastic) surface states, whose effects must be contained within the buffer region, which in practice determines the required buffer width. For insulators dangling bonds are created whose effects can be suppressed through hydrogen bond termination, whilst in metals a charge dipole is induced with decaying Friedel oscilations~\cite{ashcroft}. As the buffer region is treated in DFT only to provide correct forces in the QM region, forces on atoms in the buffer (and bulk) region are given by the classical force field, following the `abrupt force mixing' coupling scheme, which gives accurate forces throughout the overall QM/MM system, in contrast to other handshaking methods that typically incur large force errors close to the QM/MM interface~\cite{Bernstein2009}.

Here, DFT calculations are performed on clusters composed of QM and buffer atoms surrounded by vaccuum. Alternative embedding approaches have been proposed for metals that use periodic QM calculations surrounded by bulk-like regions instead of vacuum~\cite{Woodward2005,Meyer2014,Huber2016}. For example \citet{Woodward2005} modelled dislocation cores in periodic DFT cells by incorporating a domain boundary at the edge of the cell. This is appropriate where the embedding region is bulk-like; however, the topology of dislocations and cracks of interest here restricts the general applicability of such an approach. We therefore used mixed boundary conditions, with periodicity retained in the direction along dislocation lines and crack fronts, and vacuum added in the other two directions.

For hybrid simulation schemes to produce accurate results, the quantum/classical transition region should typically be only weakly deformed by the presence of the defect, such that an interatomic potential with identical elastic properties and lattice constants $(B_{\rm cl},a_{\rm cl})$ to the DFT system $(B_{\rm qm},a_{\rm qm})$ would give an identical mechanical response. However, whilst modern interatomic potentials typically reproduce DFT elastic properties well the agreement is not perfect; as a result, the atomic positions used for the classical calculation must be scaled by a factor $\alpha = a_{\rm qm}/a_{\rm cl}$ such that atoms in a perfect bulk lattice are fully relaxed in both systems. In addition, using the classical and quantum bulk moduli $B_{\rm cl}$ and $B_{\rm qm}$ to represent the elastic properties of each medium, the classical atomic forces are scaled by a factor $\alpha\beta$, where $\beta=B_{\rm qm}/\alpha^3B_{\rm cl}$. A derivation of this scaling is given in the appendix. Beyond elastic matching, there are no further QM/MM interaction terms to be calibrated, unlike for energy-based QM/MM schemes where an interaction potential describing the energetic coupling between QM and MM regions must be specified~\cite{Zhang2010}.

Our hybrid force-mixing implementation was performed in the Atomic Simulation Environment\cite{Larsen2017}, using LAMMPS\cite{LAMMPS} to generate classical interatomic forces and VASP\cite{VASP} to perform DFT simulations using projected augmented wave pseudopotenitals\cite{PAW}. To test the force mixing scheme and buffer size we first considered a perfect fcc lattice of aluminium, using an embedded atom method (EAM) interatomic potential by Liu {\it et al.}\cite{liu2004}. The QM region was a cube of 13 atoms, with a buffer region of width $w$ containing all atoms within a distance $w$ from an atom in the QM region. In this instance the DFT system is a free cluster meaning only a $\Gamma$-point calculation is required, with a plane wave cutoff of 320eV. As there should be no residual forces on atoms in a perfect lattice configuration, we measured the total magnitude of atomic forces on all atoms in the QM region with buffer size. As shown in figure \ref{vacancy}, convergence was achieved for a buffer width of 6.5~{\rm\AA}, or around three atomic layers, with the total residual atomic force in the QM region being around 10$^{-3}$~eV/\AA{}, well below the tolerance of at most 10$^{-2}$~eV/\AA{} per atom used during structural minimisation.

\section{Virtual Work Principle}
The virtual work principle states that the energy $\Delta E(r)$ required to traverse a pathway ${\bf U}(r)\in\mathbb{R}^{\rm 3N}$, $r\in[0,1]$ in configuration space is given by
\begin{equation}
    \Delta E(r) = -\int_{{\bf U}(0)}^{{\bf U}(r)}{\rm d}{\bf X}\cdot{\bf F}({\bf X}) = -\int_0^r{\rm d}r'\frac{\partial{\bf U}}{\partial r'} \cdot {\bf F}(r'),\label{VW}
\end{equation}
where ${\bf F}(r)\equiv{\bf F}({\bf U}(r))\in\mathbb{R}^{\rm 3N}$ is the force vector for a given configuration ${\bf U}(r)$. When the force is a gradient field of some energy function $V({\bf X})$ (for which only the spatial gradient, namely the force, can be calculated in hybrid simulations) it is simple to show that $\Delta E(r)= V({\bf U}(r))-V({\bf U}(0))$. We have implemented the virtual work principle in a modified nudged elastic band constrained minimisation routine\cite{neb}, evaluating ${\bf U}(r)$ from a splined set of (possibly unconverged) NEB knots, using (\ref{VW}) to extract energy differences along the pathway. In the NEB routine an energy functional is only required to define the climbing image and, in some variations of the method, to determine the finite difference scheme used to construct pathway tangents. As a result, we first run iterations with no climbing image defined, until a certain tolerance in the maximum force component perpendicular to the pathway is reached, then use (\ref{VW}) to define energy differences along the pathway to identify a climbing image. Typically, a larger number of knots are required as compared to standard NEB calculations, to ensure the splined configuration is as smooth as possible; we have found 10-15 knots to be adequate for all the systems considered here. In figure \ref{vacancy} we demonstrate an implementation of this method for the migration of a vacancy in face centered cubic (fcc) aluminium, treated in the hybrid scheme using only a single $\Gamma$-point calculation as the DFT region is again a free cluster. We also performed the same simulations in pure DFT, using a $3\times3\times3$ supercell of 107 atoms and a $7\times7\times7$ k-point grid. In the latter case we are able to extract the total energy and therefore compare the accuracy of our method; as can be seen in figure \ref{vacancy}, it is clear that the virtual work energy landscape as calculated in hybrid and DFT and the energy landscape as extracted from the total DFT energy are in extremely good agreement, demonstrating the convergence of the hybrid scheme and the validity of the virtual work principle. In contrast to the LOTF scheme, in which the QM region can be moved during a dynamical simulation, here we use the same set of QM atoms for all knots along the NEB path.

\section{Migration of an $\langle100\rangle(010)$ Edge Dislocation In Mo}
$\langle100\rangle(010)$ edge dislocations in bcc metals are known to migrate through a double kink mechanism\cite{swinburne2013} and play an important role in irradiation damage of bcc metals, forming the core of $\langle100\rangle$ prismatic dislocation loops\cite{Dudarev2008}. Edge dislocations possess a strong, long ranged deformation field which, unlike $1/2\langle111\rangle$ screw dislocation dipoles\cite{Rodney2009,Dezerald2014} cannot be contained in periodic DFT supercells. As a result, flexible boundary DFT calculations\cite{woodward2002} or the hybrid methods presented here must be used to capture the long range elastic field. NEB calculations have been successfully applied to calculate the Peierls barrier to rigid dislocation motion in a wide variety of materials\cite{Rodney2009,Dezerald2014,groger2012,rodney2017}. However, in order to correctly calculate a Peiels stress\cite{Dezerald2014,groger2012,Dezerald2016}, care must be taken to accurately determine the dislocation core position as a function of the NEB coordinate $r$, captured through some remapping function ${\rm x}_{\rm dislo}(r)$. In the present setting the complication of finding a suitable  ${\rm x}_{\rm dislo}(r)$ does not arise as we focus on the zero stress double kink formation energy, which controls the thermally activated diffusion of $\langle100\rangle$ prismatic dislocation loops\cite{dudarev2014}. By the chain rule one can demonstrate that the maximum energy difference $\Delta E = \max_r\left[-\int_0^r{\rm d}r'\partial_{r'}{\bf U}\cdot {\bf F}(r')\right]$ obtained in the virtual work expression (\ref{VW}) is invariant under the substitution ${\rm x}_{\rm dislo}(r)$ and thus does not affect our results. We note that existing methods\cite{Dezerald2014,rodney2017} to calculate the Peierls stress through the determination of a suitable function ${\rm x}_{\rm dislo}(r)$ can be applied in post processing, without modification, to the NEB pathways produced using our approach.

\begin{figure}[!ht]
    \includegraphics{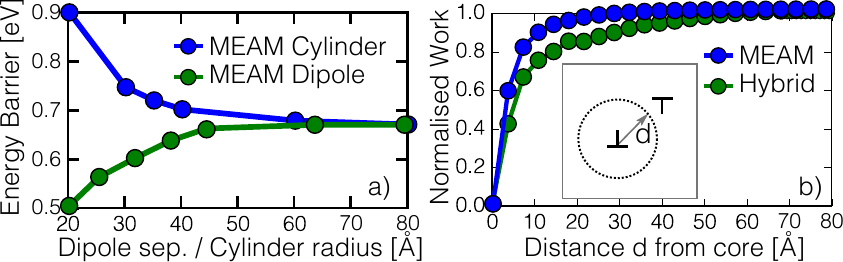}
    \caption{a) Comparison of dipole and cylinder geometries to calculate the Peierls barrier of an $\langle100\rangle(010)$ edge dislocation in MEAM molybdenum. b) Localised work for the MEAM and hybrid systems as described in the main text, using equation (\ref{VW_ind})}
    \label{mo_edge_conv}
\end{figure}

An $\langle100\rangle(010)$ edge dislocation dipole of length $b=|a[100]|$ was formed in a square supercell, such that the dislocation dipoles are separated by half the supercell height, with one dislocation migrating by $a[001]$. The system was relaxed using a recently developed modified embedded atom method (MEAM) potential by Park {\it et al.}\cite{park2012}, which includes an angular dependence to capture the highly directional bonding of bcc metals. We find the MEAM migration barrier converges with increasing system size and dipole separation, as shown in figure \ref{mo_edge_conv}; this size convergence was confirmed in calculations with a single dislocation in a cylindrical supercell, the outermost atoms fixed to the displacements predicted by anisotropic elasticity theory\cite{clouet2009}. The size convergence of the migration barrier can also be investigated by considering the {\it local} work done by an atom $j$ along the migration pathway
\begin{equation}
  W_j(r) = -\int_0^r{\rm d}r'\frac{\partial{\bf u}_j}{\partial r'} \cdot {\bf f}_j(r'),\label{VW_ind}
\end{equation}
where ${\bf u}_j,{\bf f}_j\in\mathbb{R}^3$ are the per-atom values of ${\bf U},{\bf F}$. With a saddle point $r=r_s$, the locality of the total work can be probed by summing all values of $W_j(r_s)$ less than a distance $d$ from the dislocation core, shown in figure \ref{mo_edge}b. Whilst it is clear that the immediate core region gives the dominant contribution to the migration barrier, the far field is essential to give a convergence result, which is only accessible to the hybrid simulation technique presented here. The final system used for our hybrid simulations consisted of around 10,000 atoms, far too large for a purely {\it ab initio} treatment.

\begin{figure}[!ht]
    \includegraphics[width=\columnwidth]{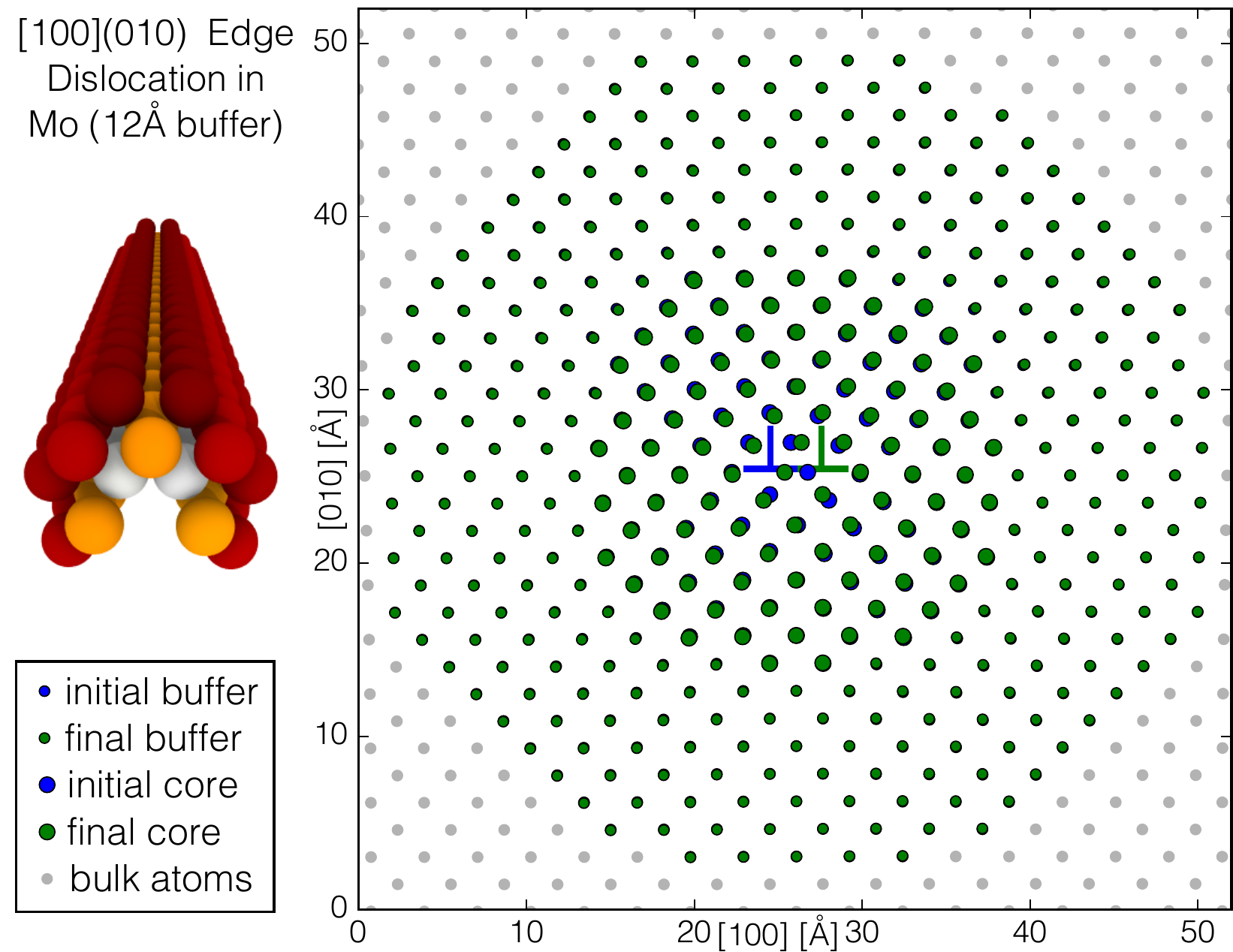}
    \caption{Illustration of the buffer and QM regions used to treat $\langle100\rangle(010)$ edge dislocations in this work. The supercell has periodic boundary conditions in the dislocation line direction, $[001]$.}
    \label{mo_edge_buffer}
\end{figure}

In the hybrid simulations, illustrated in figure (\ref{mo_edge_buffer}), the QM region was defined to contain three atomic planes around the joint initial and final positions of the moving dislocation, with the surrounding buffer region constructed as before. Although the DFT simulation has free surfaces normal to the dislocation line, the supercell remains periodic along the line direction, meaning that we must introduce k-points in one dimension\cite{Martin}. Figure \ref{mo_edge} shows the result of NEB calculations to determine the Peierls barrier of the dislocation using a variety of buffer widths and total number of k-points. Unlike the vacancy and pure bulk systems, where a buffer size of around three atomic planes was required, for the dislocation system we require a buffer of at least five atomic planes leading to DFT clusters containing around 400 atoms, which we attribute to the much greater degree of deformation caused by the dislocation and the lower atomic density of the bcc structure. Nevertheless, across the range of buffer sizes and k-points we find a variation from the final converged value of around 10\%. Significantly, the value from our hybrid simulations is around five times smaller than that found using the MEAM potential, demonstrating the importance of using {\it ab initio} forces to treat highly deformed defect cores.

\subsection{Kink Formation Energy}
\begin{figure}[b]
    \includegraphics{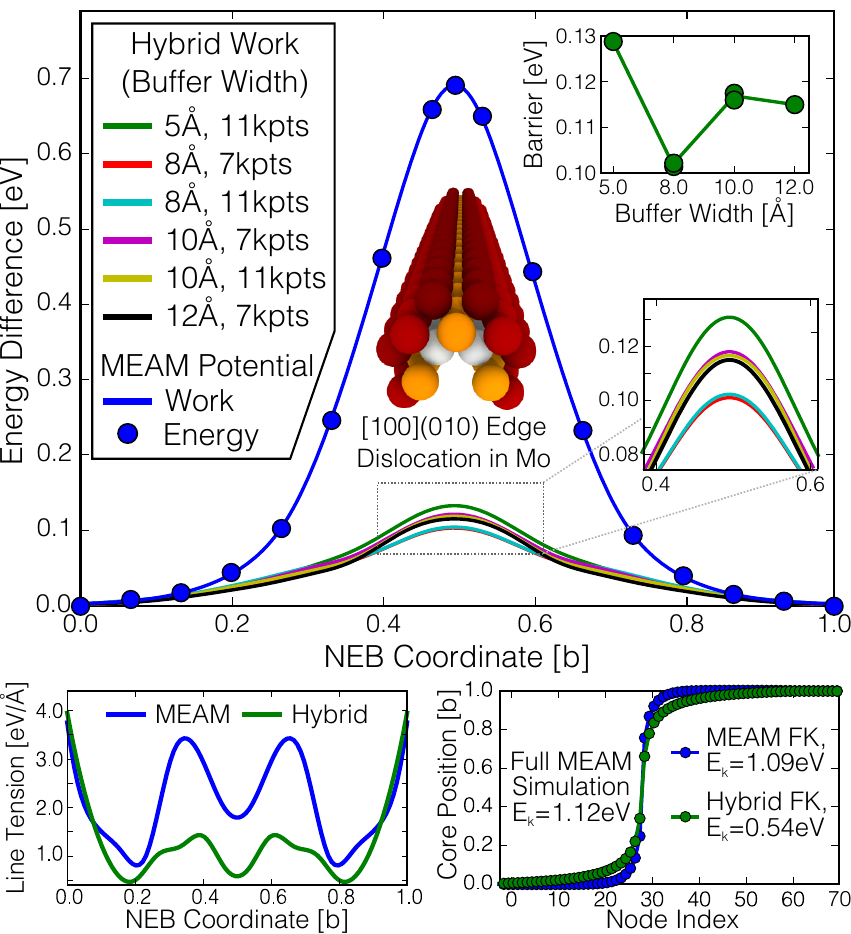}
    \caption{Top: Comparison of NEB methods to calculate the Peierls barrier of an $\langle100\rangle(010)$ edge dislocation in molybdenum. Bottom Left: Line tension calculation using hybrid and MEAM force fields as described in the text. Bottom Right: Kinks formed in Frenkel Kontorova models using the hybrid and MEAM values.}
    \label{mo_edge}
\end{figure}

As the dislocations under study migrate through a kink mechanism, we have estimated the kink formation energy through careful parametrisation of the well known Frenkel-Kontorova (FK) model\cite{swinburne2013,dezerald2015}. In the FK model, a regular array of $N$ nodes of spacing $a$ along the dislocation line direction are free to move in the perpendicular glide direction with positions $(ia,{\rm x}_i)$, $i\in[0,N-1]$. The nodes are coupled by a line tension of strength $\Gamma$ and a Peierls potential $V_{\rm P}({\rm x})=V_{\rm P}({\rm x}+b)$ of period $b$, which is given in the current setting by the Peierls potential $V(r)$ shown in figure \ref{mo_edge} using the linear relation ${\rm x}_i = {\rm x}_{\rm dislo}(r_i) = br_i$, giving $V_{\rm P}({\rm x}) = V({\rm x}/b) = V(r)$.

The FK model has been successfully applied to calculate the kink formation energy on screw dislocations in bcc metals\cite{swinburne2013,dezerald2015}, though in the current setting we have found it necessary to allow an additional position dependence in $\Gamma(r)$, giving a total FK system energy
\begin{equation}
    E_{\rm FK}[\{r\}] = \sum_{i=1}^N \frac{b^2}{a}
    \frac{\Gamma(r_i)+\Gamma(r_{i-1})}{4}
    \left(r_i-r_{i-1}\right)^2 + V(r_i),
    \label{fkm}
\end{equation}
where $V(r_i)$ is the migration potential shown in figure \ref{mo_edge}. To determine $\Gamma(r)$ and thus the kink formation energy, we will conjoin two copies of the dislocation core configurations with the same core positions $r$ and calculate the restoring force between when the core positions differ by a small quantity $\delta$. Explicitly, a line profile $r_i = r + \Theta(i-N/2+1/2)\delta$ can be formed in atomistic simulations (with $N=2$) by conjoining two NEB configurations ${\bf U}(r)\in\mathbb{R}^{\rm3N}$ and ${\bf U}(r+\delta)\in\mathbb{R}^{\rm3N}$ along the dislocation line direction to give an expanded system ${\bf U}_{\rm ext}(r,\delta)\in\mathbb{R}^{\rm6N}$, being a dislocation line twice the original length. In this extended system we can calculate the force to perturb the relative core positions by $\delta$ by projecting the force from atomistic simulations ${\bf F}[{\bf U}_{\rm ext}(r,\delta)]\in\mathbb{R}^{\rm6N}$ against the tangent$(\partial/\partial\delta){\bf U}_{\rm ext}(r,\delta)\in\mathbb{R}^{\rm6N}$, yielding a restoring force
\begin{equation}
  f(r,\delta) = \frac{\partial{\bf U}_{\rm ext}(r,\delta)}{\partial\delta}\cdot{\bf F}[{\bf U}_{\rm ext}(r,\delta)].
\end{equation}
The same line profile $r_i = r + \Theta(i-N/2+1/2)\delta$ (where again $N=2$) can also be constructed in the FK model, which yeilds a restoring force to order $\delta$ of
\begin{equation}
    f_{\rm FK}(r,\delta) = -\frac{2b^2}{a}\Gamma(r)\delta - \left(\partial_rV(r)+\delta\partial^2_rV\right) + O(\delta^2).
\end{equation}
As we have already calculated $V(r)$ through a spline interpolation we can readily calculate the derivatives $\partial_rV(r)$ and $\partial^2_rV(r)$ and thus can determine $\Gamma(r)$ by setting $f_{\rm FK}(r,\delta)= f(r,\delta)$, which yeilds
\begin{equation}
  \Gamma(r) = \frac{-a}{2b^2\delta}\left[f(r,\delta) + \partial_rV(r)\right]-\frac{a}{2b^2}\partial^2_rV + O(\delta).
\end{equation}
We emphasize that whilst to extract accurate Peierls stresses a function ${\rm x}_{\rm dislo}(r)$ which correctly extracts the `true' dislocaiton position is required, the formation energies calculated using the virtual work technique detailed here are independent of the choice of ${\rm x}_{\rm dislo}(r)$.

We have performed these calculations using both MEAM and hybrid forces to evaluate $f(r,\delta)$, yielding a calculation of $\Gamma$ which we show in lower portion of figure \ref{mo_edge}. The FK model (\ref{fkm}) can then be used to simulate a much longer dislocation line to obtain a kink formation energy, which in the pure MEAM case can be directly compared to the kink formation energy in molecular statics\cite{swinburne2013}. This technique yields a kink formation energy of 1.12~eV which is closely approximated by the MEAM FK kink formation energy of 1.09~eV. Due to the lower line tension and Peierls barrier found in our hybrid simulations, we find a much lower kink formation energy of 0.54~eV, just less than half the MEAM value. It is interesting to note that similar calculations\cite{dezerald2015} on $\langle111\rangle$ screw dislocations in bcc Mo, using periodic DFT supercells, find a kink formation energy of 0.52~eV, meaning that both slip systems have similar activation energies for plastic flow\cite{hirth}.

\section{Brittle Crack Growth in Silicon}
\begin{figure}
    \includegraphics[width=0.5\textwidth]{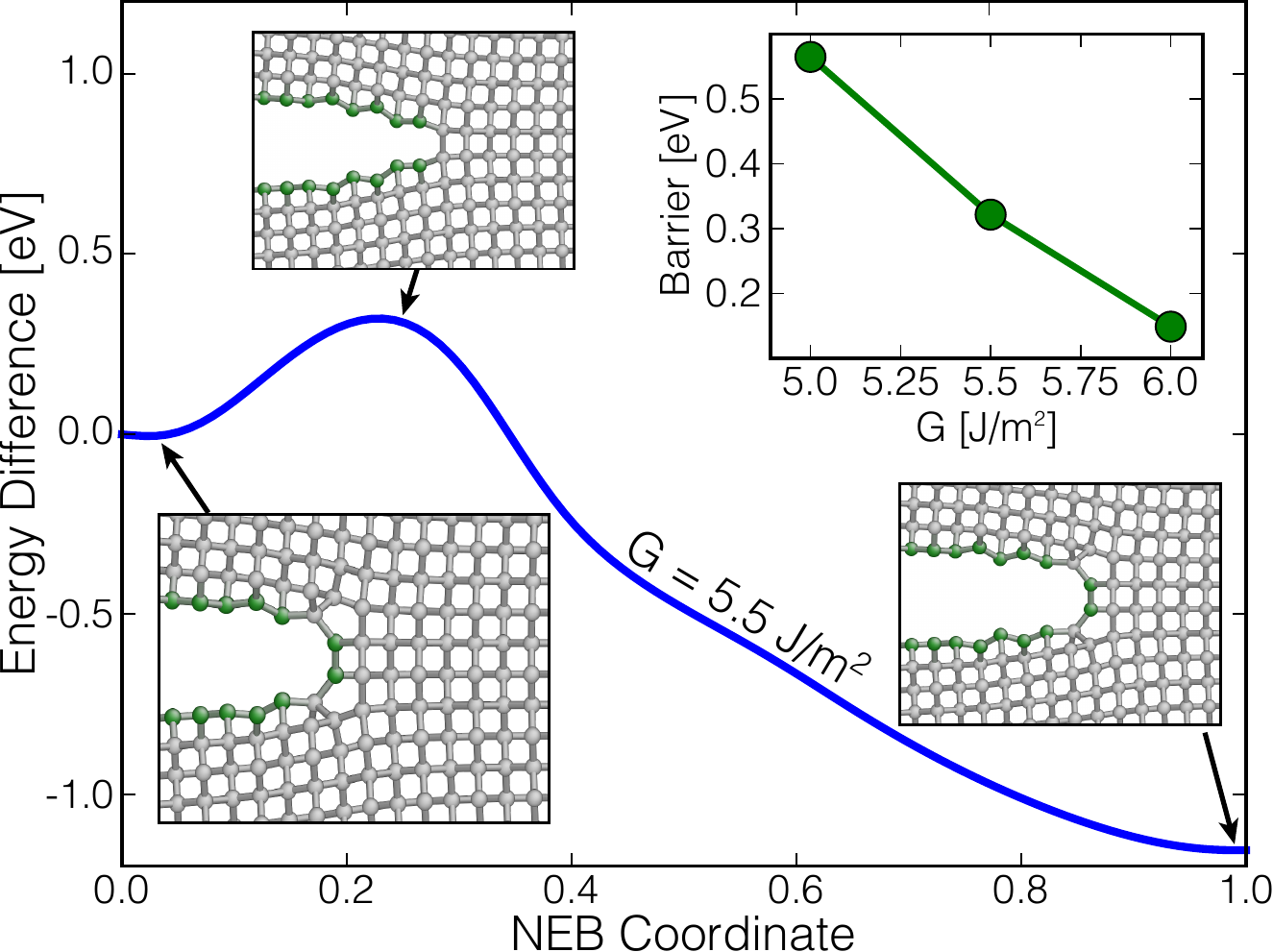}
    \caption{Minimum energy path for cleavage in the Si$(110)[1\bar{1}0]$ fracture system, involving a blunt-sharp-blunt tip reconstruction with two bonds opening simultaneously. Inset images show near-tip region for the initial, transition and final states following hybrid relaxation of the path, with under coordinated atoms shown in green. Upper right inset gives dependence of energy barrier on strain energy release rate, with fracture becoming easier as load increases.}
    \label{fig:crack}
\end{figure}

As a final example, we carry out the first direct {\em ab initio} calculation of the discrete lattice trapping barriers~\cite{Thomson1971,Bernstein2003} to brittle crack growth in silicon in the $(110)[1\bar{1}0]$ cleavage system (figure~\ref{fig:crack}).
We modelled an 8,112 atom system with dimensions $297\times97.9\times5.43$~\AA$^3$, periodic along the crack front direction and  with clamped top and bottom edges and applied strains corresponding to strain energy release rates of 5.0~J/m$^2$, 5.5~J/m$^2$ and 6.0~J/m$^2$ (above the Griffith load of 3.44~J/m$^2$ computed from the relaxed DFT surface energy~\cite{Kermode2015}).
An initial configuration with $N = 70$ broken bonds along the crack line was relaxed using the force-based hybrid scheme using the Stillinger-Weber potential~\cite{Stillinger1985} for the MM region and DFT with the PBE exchange-correlation functional for a QM region containing 32 atoms centred on the crack tip plus a buffer radius of 6~\AA{}. The corrugated reconstruction of the $(110)$ surface leads to a slightly blunted crack tip, with an alternating up-down structure that means the next stable minimum occurs with $N+2$ broken bonds. Applying the virual work NEB approach with $(N, N+2)$ end points identifies a minimum energy path where two diagonally oriented bonds cleave simultaneously, passing through a sharp-tip transition state. The lattice trapping barrier decreases as the strain energy release rate is increased, predicting thermally activated crack growth rates similar to earlier work where DFT barriers could only be estimated from cluster calculations~\cite{Kermode2015}, but with the tip-blunting reconstruction indicating slow crack growth remains important for larger strain energy release rates than previously thought.

\section{Conclusion}
In summary, we have proposed a method to compute energy barriers for activated processes that combines DFT and classical interatomic potentials in materials systems where strong bonds cross the interface between QM and MM regions. The method has been used to perform the first {\em ab initio} calculation of the Peierls barrier for $\langle100\rangle(010)$ edge dislocations in Mo and to identify a novel crack advance mechanism in Si. The method is expected to be generally applicable to any system where localised chemical processes are driven by long range elastic fields. For example, the technique could be applied to provide {\em ab initio} mechanistic insight into the dynamics of three-dimensional crack fronts, where fracture proceeds through kink formation and advance~\cite{Kermode2015}, or to provide a QM-based analogue of the Rice-Thomson criterion for the transition from brittle cleavage to dislocation emission~\cite{Rice1974,Andric2017}.
\section{Acknowledgements}
This work has been carried out within the framework of the EUROfusion Consortium and has received funding from the Euratom research and training programme 2014-2018 under grant agreement No 633053, from the RCUK Energy Programme [grant number EP/I501045], and from the Engineering and Physical Sciences Research Council under grant numbers EP/L027682/1 and EP/P002188/1. Computing facilities were provided by the Scientific Computing Research Technology Platform of the University of Warwick and the EUROfusion Marconi supercomputer facility. The views and opinions expressed herein do not necessarily reflect those of the European Commission.
\appendix
\section{Derivation of scaling laws}
We wish to define a position scaling $\alpha$ and energy scaling $\beta$ on the classical system to match the bulk moduli and lattice constant of the quantum system. We define a potential energy function $E({\bf X})$, and then a
scaled function \[ E'({\bf X}) = \beta E(\alpha {\bf X}) \]
The corresponding force in the original coordinate system is
\[
{\bf F}'({\bf X}) = -\frac{\partial E'}{\partial {\bf X}} = -\beta \alpha \frac{\partial E'}{\partial {\bf X}} = \beta \alpha\, {\bf F} (\alpha {\bf X})
\]
The equilibrium lattice constant changes from $a_0$ to $a_0'$ and the
equilbrium cell volume changes from $V_0$ to $V_0'$ according to
\[
a_0' = \frac{a_0}{\alpha},\quad
V_0' = \frac{V_0}{\alpha^3}.
\]
The scaled bulk modulus is
\[
B' = V \frac{\partial^2 E'}{\partial V^2} \quad
  = \beta \alpha^3 V \frac{\partial^2E}{\partial V^2}  \quad
  = \beta \alpha^3 B
\]%
Thus if we want to match a target volume $V_0'$ and bulk modulus $B'$
we should use
\[
 \alpha = \left( \frac{V_0}{V_0'} \right)^\frac{1}{3} = \frac{a_0}{a_0'},\quad
  \beta  = \frac{B'}{B \alpha^3}
\]%
where $a_0$ and $a_0'$ are the lattice constants before and after
rescaling. For quantum / classical force mixing, where we label the quantum region as 1 and the classical region as 2, the aim is to rescale
the classical region to match the quantum lattice constant $a_1$ and bulk
modulus $B_1$, so we have
\[
  \alpha = \frac{a_1}{a_2},\quad
  \beta  = \frac{B_1}{B_2 \alpha^3}
\]
where $a_2$ and $B_2$ are the unmodified classical lattice constant and bulk
modulus, respectively. The force scaling is thus $\alpha\beta = B_1/\alpha^2B_2$ as given in the main text.\\


%

\end{document}